%% file: UDFhighz.tex
\documentclass[12pt,preprint]{aastex}

\begin{document}

\title{Candidates of $z\simeq$ 5.5--7 Galaxies in the HST Ultra Deep Field }

\author{Haojing Yan\altaffilmark{1} \& Rogier A. Windhorst\altaffilmark{2}}

\altaffiltext{1} {Spitzer Science Center, California Institute of Technology,
MS 100-22, Pasadena, CA 91125; yhj@ipac.caltech.edu}
\altaffiltext{2} {Department of Physics \& Astronomy, Arizona State University, Tempe, AZ 85287}

\begin{abstract}

  We report results from our $z\simeq 5.5$--7 galaxy search in the HST Ultra
Deep Field (UDF). Using the 400-orbit of ACS data, we found 108 plausible
$5.5\leq z\leq 6.5$ (or $z\simeq 6$ for short) candidates to
$m_{AB}(z_{850})=30.0$ mag. The contamination to the sample, either due to
image artifacts or known types of astronomical objects, is likely negligible.
The inferred surface densities of $z\simeq 6$ galaxies are consistent with our
earlier predictions from $m_{AB}(z_{850})=26.5$ to 28.5 mag. After
correcting for detection incompleteness, the counts of $z\simeq 6$ candidates
to $m_{AB}(z_{850})=29.2$ mag suggests that the faint-end slope of the galaxy
luminosity function (LF) at this redshift is likely between $\alpha=-1.8$ and
$-1.9$, which is sufficient to account for the entire Lyman photon budget
necessary to complete the reionization of the universe at $z\simeq 6$. We also
searched for $z\simeq 6.5$--7 candidates using the UDF NICMOS data, and have
found four candidates to $J_{110}=27.2$ mag. However, the infrared colors of
three candidates cannot be easily explained by galaxies in this redshift range.
We tentatively derive an upper limit to the cumulative surface density of
galaxies at $z\simeq 7$ of 0.36 per arcmin$^2$ to $J_{110}=26.6$ mag, which
suggest a noticeable drop in the LF amplitude from $z\simeq 6$ to $z\simeq 7$.

\end{abstract}


\section{Introduction}

   Using the Advanced Camera for Surveys (ACS), a public, ultra-deep survey has
been carried out by the Hubble Space Telescope (HST). This Ultra Deep Field
(UDF; PI. S. Beckwith) campaign observed a single ACS Wide Field Camera (WFC) 
field within the Chandra Deep Field South in four broad-bands covering 0.4 to
nearly $1.0\mu m$. To enhance the value of these ACS data, the Camera 3 (NIC3)
of the Near Infrared Camera and Multi Object Spectrometer (NICMOS) has observed 
the central portion of this field in the F110W ($J_{110}$) and the F160W 
($H_{160}$) filters (PI. R. Thompson). With a total exposure of 400 orbits in
the ACS and 144 orbits in the NICMOS, the UDF will remain the deepest
optical/IR survey field in the coming seven years. Here we discuss the 
$z\simeq 5.5$--7 candidates found in this field. We adopt the following 
cosmological parameters: $\Omega_M=0.27$, $\Omega_\Lambda=0.73$, and 
$H_0=71$ km$\,$s$^{-1}\,$Mpc$^{-1}$. All magnitudes are in AB system.

\section{Data and Selection of $z\simeq 6$ and $z\simeq 7$ Candidates}

   The total UDF ACS/WFC exposure times are 37.5, 37.6, 96.4,
and 96.3 hours in the F435W ($B_{435}$), F606W ($V_{606}$), F775W ($i_{775}$),
and F850LP ($z_{850}$) filters, respectively. The final drizzle-combined stacks
have a pixel scale of 0.03$^{''}$, and cover an effective area of 10.34 
arcmin$^2$ after trimming off the lower S/N edges. We performed matched-aperture
photometry independently using SExtractor (Bertin \& Arnouts 1996) in
double-input mode with the $z_{850}$ stack as the detection image. We used a
$5\times 5$ Gaussian convolving kernel with a FWHM of 3 pixels, and required
a real detection have a minimum of 5 connected pixels 1.5 $\sigma$ above 
background. The NIC3 UDF program observed a $3\times 3$ grid that covers the
center of the UDF, giving an effective coverage of 5.76 arcmin$^2$ with an 
average exposure time of 5.97 hours in both the $J_{110}$ and the $H_{160}$
filters. The final drizzle-combined stacks have a pixel
scale of 0.09$''$. We used the photometric catalog that comes with
the data release, which was generated based on the detections in the co-added 
$J_{110} + H_{160}$ stack using SExtractor. The ``MAG\_AUTO'' option was used
in both cases.
 
   We selected $z\simeq 6$ candidates as $i_{775}$ drop-outs in the UDF ACS
images. Instead of aiming at $z\geq 6.0$ and using the color criterion of 
($i_{775}-z_{850}$) $\geq 2.0$ mag as in our previous work
(Yan, Windhorst \& Cohen 2003), we here target at $5.5\leq z\leq 6.5$ and
adopt ($i_{775}-z_{850}$) $\geq 1.3$ mag as the first criterion. In total, 108
objects were selected by using this color criterion alone. These objects were
visually examined in all the four bands to ensure that there was no obvious
reason (e.g., image defects) to exclude them from the candidate list. The
second criterion is that a valid candidate should not be detected in $B_{435}$
and $V_{606}$, {\it i.e.}, it should either have reported magnitude fainter 
than 29.5 mag, or have its estimated photometric error larger than 0.54 mag 
($S/N<2$). All of the 108 candidates satisfy this criterion. The coordinates
and the photometric properties of all the 108 candidates are listed in Table 1.
These objects seem to be strongly clustered, e.g., we identified 6 multiple
systems whose members are within $1^{''}$ (or $<5.8$ co-moving kpc) from each
other. The significant fraction of multiple systems might indicate that 
merging was rather common at $z\simeq 6$. 

   Fifty two of the 108 candidates are within the NIC3 mosaic, among which 12
objects (all at $z_{850}<27.8$ mag) have NIC3 counterparts.
Actually, 9 of these 12 objects are in 4 multiple systems, and the
coarser pixel resolution of the NIC3 images cannot resolve their individual
members and thus identify them as single sources. The
$z_{850}J_{110}H_{160}$ colors of these 12 objects
(9 sources in 4 multiple systems, and 3 isolated sources) are shown in Fig. 1.
The $z_{850}$ magnitudes of the multiple systems are derived by adding the
fluxes of their individual members. The known $z=5.83$ galaxy 
(Dickinson et al. 2004) is among these 4 multiple systems (ID 1a in Table 1).
   
   Candidates of $z\simeq 7$ objects were selected as $z_{850}$-band drop-outs.
As the $J_{110}$ band heavily overlaps with the $z_{850}$ band, the 
($z_{850}-J_{110}$) color at $z>6.5$ does not change as dramatically
as the ($i_{775}-z_{850}$) color does at $z\simeq 6$. We adopted the criteria
as ($z_{850}-J_{110}$) $>0.8$ mag and ($J_{110}-H_{160}$) $>-0.1$ mag,
and no detection in ACS $B_{435}$, $V_{606}$ and $i_{775}$ bands.
This search resulted in one $z_{850}$ source and three $z_{850}$ drop-outs, 
all of which were visually inspected and deemed reliable. 
The coordinates and the photometric properties of the three $z_{850}$ drop-outs
are listed in Table 2. The one $z_{850}$ source turns out to be the multiple
system 7a/7b in Table 1 ({\it i.e.}, also qualifies as being $z\simeq 6$), and
thus is not listed again.

\section{Discussion of the $z\simeq 6$ Candidate Sample}

\subsection{Consistency of the $z\simeq 6$ Interpretation}

   It is known that brown dwarfs can mimic the broad-band colors of a
$z\simeq 6$ galaxy because of their strong molecular absorption bands.
The 4000\AA\ break in elliptical galaxies at $z\simeq 1.0$--1.5 can also mimic
the Lyman-break at $z\simeq 6$ (e.g., Yan, Windhorst \& Cohen 2003). Lower-z
late-type galaxies, even those significantly reddened by dust, generally are
not a significant interloper, because their 4000\AA\ breaks are not as strong
as those in the early-type galaxies.

   The contamination due to brown dwarfs is insignificant in our case, as only
4 of our 108 candidates have SExtractor star/galaxy separation flag larger than
0.90. This flag is 0 for extended sources and 1 for point sources. The colors
of typical M-, L-, and T-type brown dwarfs (e.g., Kirkpatrick et al. 1999) are
shown in Fig. 1. None of the $z\simeq 6$ candidates with NIC3 measurements has
IR colors close to the loci of brown dwarfs. The contamination due to 
elliptical galaxies at $z\simeq$1.0--1.5 is likely also small. While the major
color criterion is ($i_{775}-z_{850}$) $>$ 1.3 mag, most of the candidates have
($i_{775}-z_{850}$) $>$ 1.5 mag (90 out of 108), and thus the chance for a 
low-z elliptical galaxy to be selected as candidate is greatly reduced, since
the latter usually have ($i_{775}-z_{850}$) $\leq 1.3$ mag. Further evidence
comes from the candidates also detected in NIC3 images. Fig. 1 shows the 
``color track'' of a typical E/S0 galaxy in $z_{850}J_{110}H_{160}$ space, 
together with the tracks of a Sbc galaxy with and without dust reddening (all
templates are from Coleman, Wu \& Weedman 1980, and the extinction law is from
Calzetti et al. 2000). None of the NIC3 detected $z\simeq 6$ candidates has 
colors close to these low-z tracks. On the other hand, the colors of our 
candidates are consistent with being $z\simeq 6$, if we consider the systematic
photometric errors due to the aperture mismatching between the ACS and the NIC3
photometry, and, more importantly, the variation in the SED of $z\simeq 6$
galaxies. While a detailed stellar population synthesis approach is beyond the
scope of this Letter, we point out that most of the candidates in Fig. 1 can be
well explained by the models of Bruzual \& Charlot (2003). For simplicity, 
Fig. 1 shows the color tracks of the Simple Stellar Population (SSP) models 
with ages of 10 Myr and 300 Myr, which bracket most of the $z\simeq 6$
candidates.

\subsection{Photometric Contamination and Correction for Incompleteness}

   When the search is pushed to very faint levels, contamination due to noise
spikes could become severe. To assess the effect due to spurious detections, we
performed the ``negative source'' check as described by Dickinson et al.
(2004). The UDF ACS mosaics were inverted, and SExtractor was run on these 
inverted images using the same parameter settings as in \S 2. We find only 10 
``negative objects'' that have $S/N\geq 3$, and {\it none} of them satisfy our
color selection criterion of $(i_{775}-z_{850})\geq 1.3$ mag. This is not
surprising, because the UDF ACS mosaics were created by stacking a very large
number of dithered images (the $z_{850}$ band mosaic has 288 ditherings), so
that the image artifacts are minimal. Therefore, we conclude that spurious
detection has a negligible impact to our candidate sample.

   Another effect that should be considered is sample incompleteness, which
starts to be significant at $z_{850}>28.5$ mag. We note that the $z_{850}$-band
count histogram peaks at 28.5 mag ($S/N\simeq 10$), and drops to 50\% of this
peak value at 29.2 mag ($S/N\simeq 7$). For our purpose below, it is sufficient
to discuss only the incompleteness of the candidates in the intermediate
brightness range of 28.5$<z_{850}<$29.2 mag. We estimated the incompleteness
as following. For each of the 28 candidates in this regime, a $11\times11$ pixel
image ``stamp'' centering on the object was copied from the $z_{850}$-band
mosaic. Each ``stamp'' was then added to the $z_{850}$-band mosaic at 
$\sim 550$ randomly distributed positions. Source detection was performed on
these simulated images to recover the artificially added objects. We find that
the median recovering rate is 28.5\%, corresponding to an incompleteness
correction factor of 3.51.

\subsection{LF of galaxies at $z\simeq 6$ and Reionization}

   In Yan et al. (2002), we made a prediction of the LF of galaxies at
$z\simeq 6$ based on the measured galaxy LF at $z\simeq 3$. As summarized in
Yan \& Windhorst (2004; YW04), this LF estimate agrees well with all available
observations (e.g., Rhoads et al. (2003); Yan, Windhorst \& Cohen (2003);
Stanway, Bunker \& McMahon (2003); Bouwens et al. (2003); Dickinson et al.
(2004)). Based on this LF, it was also suggested in that paper that ``normal''
galaxies can account for the entire ionizing photon budget necessary to finish
the reionization of the universe by $z\simeq 6$, as long as the faint-end slope
of the LF is sufficiently steep. This LF, with slopes of $\alpha=-1.6$, $-1.8$
and $-2.0$, is reproduced in Fig. 2. YW04 predicted that the
UDF data would reveal 50--80 $z\simeq 6$ objects to $z_{850}=28.4$ mag. Among
the 108 candidates reported here, 55 objects are brighter than this level,
and this agrees with our earlier prediction.

   As discussed in YW04, if the nominal clumping factor value ($C=30$ at $z=5$)
is adopted, normal galaxies can account for the entire reionizing photon budget
at $z\simeq 6$ as long as the faint-end slope of the LF at this redshift is
somewhat steeper than $-1.6$ and the normalization of the LF is close to what
we estimated. If the faint-end slope is shallower than $-1.6$, galaxies cannot
meet the reionization requirement unless the clumping factor is significantly
lower, or the LF normalization is much higher, or the Lyman photon escaping
fraction is larger. Since every evidence indicates that the LF normalization is
about right, the important issue is the faint-end slope of the $z\simeq 6$ LF.
No other data set is better than the UDF $z\simeq 6$ candidate sample for this
purpose. The cumulative number densities inferred from our sample, without
incompleteness correction, are plotted as solid red boxes in Fig. 2. Again,
these numbers agree well with our LF prediction to 28.5 mag, where the
candidate selection does not suffer from severe incompleteness. However, as
Fig. 2 also shows, we have to measure the counts to significantly fainter than
28.5 mag in order to constrain the possible range of faint-end slopes. While
our data points beyond 28.5 mag seem to suggest a faint-end slope of 
$\alpha=-1.6$ or even shallower, this is rather biased by the incompleteness at
the faint levels. If we apply the correction as mentioned in \S 3.2, a steeper
slope is indicated. The open red box in Fig. 2 shows the corrected cumulative
surface density to 29.2 mag, if a factor of 3.51 is applied to the density of
the intermediate brightness group ($z_{850}=$28.5--29.2 mag), which suggests a
faint-end slope between $\alpha=-1.8$ and $-1.9$. For comparison, if the
incompleteness correction were only a factor of 2 for the intermediate 
brightness group --- contrary to what our simulation indicates --- a faint-end
slope of $-1.7$ would still be required. Therefore, we conclude that the major
sources of the reionization at $z\simeq 6$ are indeed normal galaxies with 
dwarf-like luminosities. It is interesting to note that Stiavelli et al. (2004)
have also reached the similar conclusion, although from a different approach,
that regular galaxies at $z\simeq 6$ are sufficient for reionization.

\section{Constraint to the LF at $z\simeq 7$}

    While our $z\simeq 7$ candidate sample has four objects, further 
investigation show that a $z\simeq 7$ interpretation is not straightforward.
First of all, the three $z_{850}$ drop-outs have 
($J_{110}$-$H_{160}$) $> 1$ mag, which makes them also $J_{110}$ drop-outs.
As a $z\simeq 7$ object should have similar fluxes in both $J_{110}$ and 
$H_{160}$ bands, they are unlikely to be at this redshift. We also
explored a large variety of dust-reddened young galaxy templates, and found
that none of them could have such a large break across the $J_{110}$ band. We
note that the first two objects in Table 2 are similar to the J-band drop-out 
object (but slightly brighter) found by Dickinson et al. (2000) in the HDF-N,
whose nature is yet unclear. Secondly, the remaining candidate in the 
$z\simeq 7$ sample is likely at the border of the $z\simeq 6$ bin and the
$z\simeq 7$ bin, as this candidate is actually the multiple system 7a/7b in the
$z\simeq 6$ sample. Therefore, whether this system should be counted as 
$z\simeq 6$ or $z\simeq 7$ is very uncertain at this stage.

  Nevertheless, we can still derive a useful {\it upper} limit of the cumulative
surface density of $z\simeq 7$ galaxies based on objects 7a/7b, assuming that
both members are at $z>6.5$. This limit is 0.36 per arcmin$^2$ to 
$J_{110}=26.6$ mag. While there are a number of reported Ly$\alpha$ emitters at
$z>6.5$, a direct comparison against these results is difficult, because these
Ly$\alpha$ emitters either are gravitationally lensed by foreground clusters
(Hu et al. 2002; Kneib et al. 2004) or do not have continuum magnitudes 
available (Kodaira et al. 2003; Rhoads et al. 2004). However, none of these
results seems to be in conflict with our derived upper limit.

    Assuming no evolution from $z\simeq 6$ to 7, our LF predicts that the
cumulative surface density of $z\simeq 7$ galaxies to 26.6 mag is 0.51 per
arcmin$^2$, which is somewhat higher than the observed upper limit. Thus our
data suggest a noticeable drop of the LF amplitude over the 0.16 Gyr from 
$z\simeq 6$ to 7, which we tentatively identified with the onset of galaxy 
formation and the onset of the first regular IMF of Pop II stars, the low-mass
end of which we still see in galaxy halos today, and the high mass end of which
finished reionization at $z\simeq 6$, when those hot stars resided in dwarf 
galaxies at $z\simeq 7$--6.

\section{Summary}

   We searched for $z\geq 5.5$--7 galaxy candidates in the UDF, using the 
UDF WFC and NIC3 data. We have found 108 $z\simeq 6$ candidates to a limit of
$z_{850}=30.0$ mag, which is consistent with the prediction in YW04.
The cumulative surface densities after the correction of incompleteness 
suggests a slope of $\alpha=-1.8$ to $-1.9$, which means galaxies
can account for the reionizing photon budget at $z\simeq 6$. We also
searched for $z\simeq 7$ galaxy candidates, but only found one such object 
whose redshift might be at the lower end of the redshift range under question.
The search also resulted in three $z_{850}$ drop-outs, whose colors shown that
they are not likely at $z>7$. The paucity of $z\simeq 7$ candidates
suggests the LF-amplitude drops significantly beyond $z\simeq 6$, which can be 
identified with the dawn of galaxy formation.

\acknowledgments

    The authors thank the referee for the very helpful comments. We acknowledge
the support from the NASA grants HST-GO-09780.*. 
HY acknowledges the support provided by NASA through Contract Number 1224666
issued by the Jet Propulsion Laboratory, California Institute of Technology
under NASA contract 1407.

\clearpage
\begin{figure}
\epsscale{.90}
\plotone{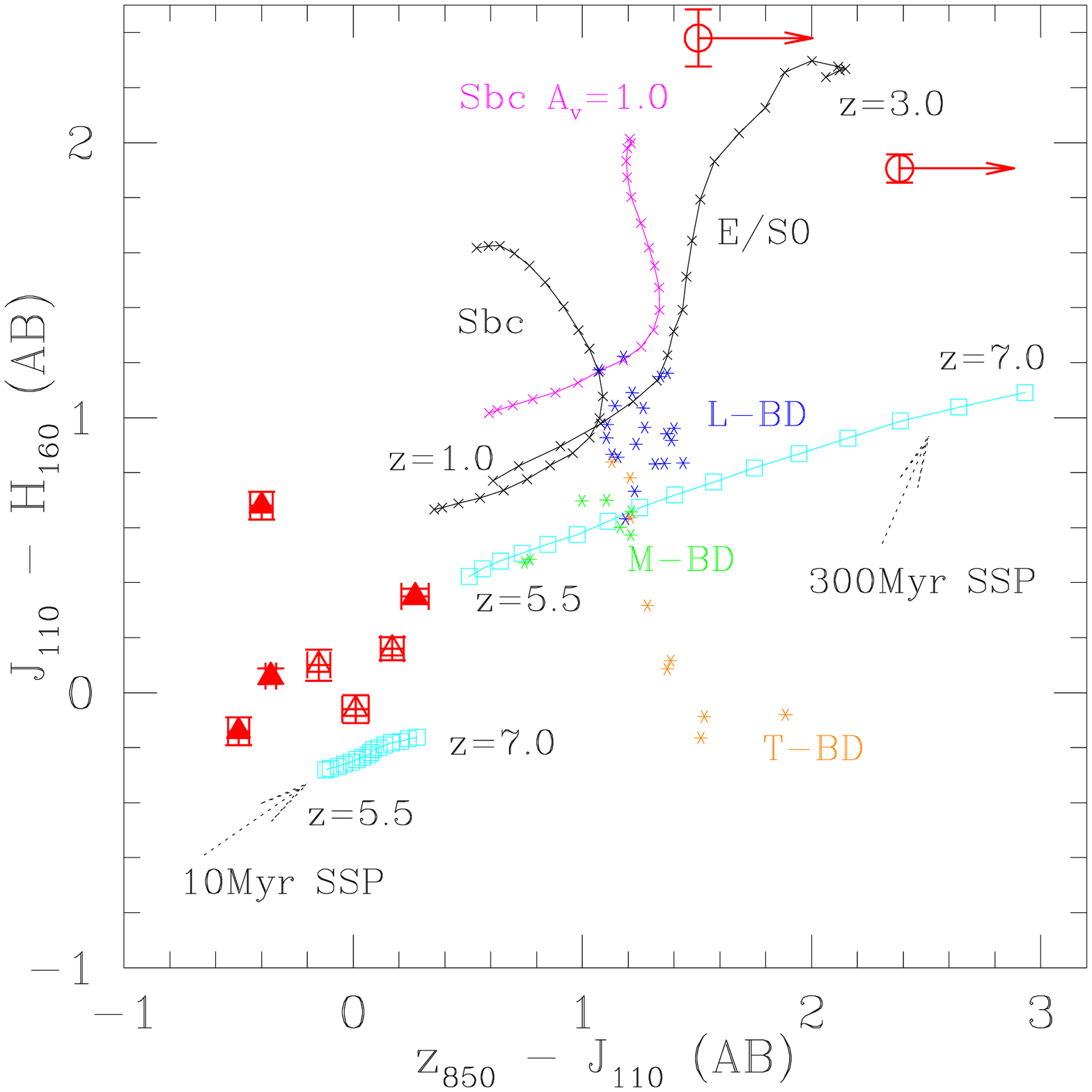}
\caption{The $z\simeq 6$ candidates detected by NIC3 are shown as red triangles
with error bars (the 4 multiple systems are shown as filled triangles). These
objects are well separated from the locations possible interlopers such as
low-z E/S0 galaxies or brown dwarfs. Their colors are consistent with 
star-forming galaxies at $z\simeq 6$, if the variations in SED (such as ages
and different star-formation history) and the systematic errors in photometry
are considered. The open circles with error-bars and $(z_{850}-J_{110})$ color
upper limits are two of the three $z_{850}$ drop-outs in the $z\simeq 7$ 
candidate sample. Their colors suggested they are not likely at high-z but more
likely lower redshift early-type galaxies.
}
\end{figure}
\begin{figure}
\epsscale{.80}
\plotone{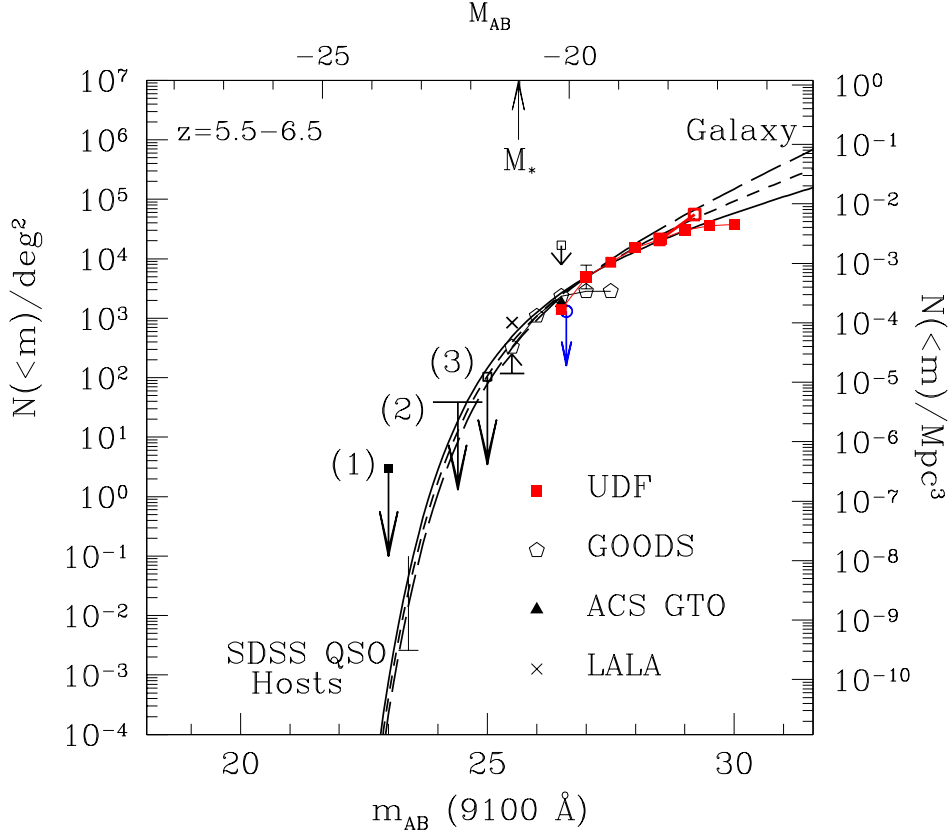}
\caption{
The cumulative surface densities of $z\simeq 6$ galaxies at different
brightness levels inferred from the $z\simeq 6$ candidates sample in this Letter
agree very well with our earlier predictions, which was extrapolated from the
galaxy LF measured at $z\simeq 3$ (Yan et al. 2002). The normalization of the LF
estimate at $z\simeq 6$ is fixed by the cumulative number density of $z\simeq 6$
galaxies in the HDF-N to a limit of $AB=27.0$ mag. This prediction is consistent
with the available observations (see YW04),
and is reproduced here for three different LF faint-end slopes: $\alpha=-1.6$
(solid line), $-1.8$ (short dashed line), and $-2.0$ (long dashed line). 
Without correction for incompleteness at faint fluxes, our UDF result suggests
a LF faint-end slope of $\alpha=-1.6$ or slightly flatter. However, after
applying a correction to incompleteness in the flux range $z_{850}=28.5$--29.2
mag, the inferred cumulative number density to 29.2 mag could be a factor of
3.51 higher (indicated by the red open box), and is consistent with a steeper
faint-end slope of $\alpha=-1.8$ to $-1.9$. Given that the incompleteness
correction cannot be zero, a faint-end slope steeper than $-1.6$ is plausible.
Based on the discussion of YW04, this clearly suggests that 
``normal'' galaxies can account for the entire reionizing photon budget at
$z\simeq 6$. The amplitude of the LF at $z\simeq 7$ could be significantly
lower, which is indicated by our derived upper limit (the blue downward arrow).}
\end{figure}

\clearpage

\input{UDFhighz_table1.tex}
\begin{table}
\caption{Photometric properties of the three $z_{850}$ drop-outs found in the UDF NICMOS field}
\begin{center}
\begin{tabular}{lcccc}\tableline \tableline

   ID &      RA\& DEC(J2000)   &  $z_{850}$(limit)$^1$  & $J_{110}$ & $H_{160}$ \\ \tableline

1 & 3:32:38.74 -27:48:39.97 & 28.967 & 26.583$\pm$0.050 & 24.677$\pm$0.012 \\
2 & 3:32:42.88 -27:48:09.52 & 28.517 & 27.012$\pm$0.103 & 24.632$\pm$0.015 \\
3 & 3:32:42.56 -27:46:56.69 & 28.449 & 27.297$\pm$0.056 & 26.106$\pm$0.024 \\

\tableline
\end{tabular}
\end{center}
\tablenotetext{1.} {These magnitude limits are obtained by adding the flux
within an aperture of 0.54$^{''}$ radius, which is not necessarily the size of 
the apertures that used for $J_{110}$ and $H_{160}$ photometry.}
\end{table}

\vfill\eject

\end{document}

%% file: UDFhighz_table1.tex
%
 
\begin{deluxetable}{cccrccccc}
\tablewidth{0pc}
\tabletypesize{\scriptsize}
\tablecaption{Photometric properties of the $z\simeq 6$ candidates}
\tablehead{
\colhead{ID} &
\colhead{RA \& DEC(J2000)} &
\colhead{S/G} &
\colhead{FWHM} &
\colhead{$i_{775}$} &
\colhead{$z_{850}$} &
\colhead{$i-z$} &
\colhead{$J_{110}$} &
\colhead{$H_{160}$}
}
\startdata

{\bf 1a} & 3:32:40.01 -27:48:15.01 & 0.04 &  5.0 & 26.88$\pm$0.03 & 25.25$\pm$0.01 & 1.62 & 25.47$\pm$0.02 & 25.41$\pm$0.02 \\
{\bf 1b} & 3:32:40.04 -27:48:14.54 & 0.00 & 12.6 & 29.03$\pm$0.18 & 27.41$\pm$0.07 & 1.62 & & \\
{\bf 2a} & 3:32:36.47 -27:46:41.45 & 0.02 &  9.6 & 29.00$\pm$0.16 & 26.49$\pm$0.03 & 2.51 & 26.60$\pm$0.04 & 25.92$\pm$0.03 \\
{\bf 2b} & 3:32:36.49 -27:46:41.38 & 0.02 &  6.1 & 29.58$\pm$0.22 & 27.76$\pm$0.07 & 1.82 & & \\
3 & 3:32:38.28 -27:46:17.22 & 0.06 &  6.0 & 29.77$\pm$0.29 & 26.68$\pm$0.03 & 3.09 & 26.83$\pm$0.04 & 26.73$\pm$0.04 \\
4 & 3:32:34.55 -27:47:55.97 & 0.02 &  6.6 & 28.62$\pm$0.10 & 26.93$\pm$0.04 & 1.70 & 26.76$\pm$0.03 & 26.60$\pm$0.03 \\
{\bf 5a} & 3:32:34.29 -27:47:52.80 & 0.00 & 14.3 & 29.07$\pm$0.21 & 26.97$\pm$0.05 & 2.10 & 26.56$\pm$0.03 & 26.70$\pm$0.04 \\
{\bf 5b} & 3:32:34.28 -27:47:52.26 & 0.00 & 14.9 & 28.98$\pm$0.19 & 27.17$\pm$0.06 & 1.80 & & \\
{\bf 5c} & 3:32:34.31 -27:47:53.56 & 0.00 & 15.3 & 29.42$\pm$0.22 & 27.76$\pm$0.08 & 1.66 & & \\
6 & 3:32:33.43 -27:47:44.88 & 0.00 &  7.5 & 29.31$\pm$0.18 & 27.23$\pm$0.05 & 2.09 & 27.22$\pm$0.03 & 27.28$\pm$0.04 \\
{\bf 7a} & 3:32:37.46 -27:46:32.81 & 0.00 & 13.0 & 32.49$\pm$4.00 & 27.50$\pm$0.07 & 4.99 & 26.61$\pm$0.02 & 26.26$\pm$0.02 \\
{\bf 7b} & 3:32:37.48 -27:46:32.45 & 0.00 & 11.7 & 29.70$\pm$0.30 & 27.78$\pm$0.09 & 1.92 & & \\
\tableline
8 & 3:32:38.02 -27:49:08.36 & 0.93 &  3.2 & 26.92$\pm$ 0.02 &  25.41$\pm$ 0.01 & 1.50  & & \\
9 & 3:32:38.80 -27:49:53.65 & 0.23 &  3.4 & 29.08$\pm$ 0.46 &  25.43$\pm$ 0.03 & 3.65  & & \\
10 & 3:32:34.09 -27:46:47.21 & 0.02 &  5.9 & 28.81$\pm$ 0.11 &  26.64$\pm$ 0.03 & 2.16  & & \\
11 & 3:32:32.61 -27:47:53.99 & 0.01 & 12.3 & 28.03$\pm$ 0.07 &  26.68$\pm$ 0.03 & 1.35  & & \\
12 & 3:32:29.98 -27:47:02.87 & 0.23 &  6.8 & 28.12$\pm$ 0.12 &  26.80$\pm$ 0.06 & 1.32  & & \\
13 & 3:32:47.85 -27:47:46.36 & 0.00 & 17.6 & 30.39$\pm$ 0.71 &  26.97$\pm$ 0.05 & 3.43  & & \\
14 & 3:32:41.57 -27:47:44.23 & 0.00 & 12.5 & 29.54$\pm$ 0.31 &  26.97$\pm$ 0.05 & 2.58  & & \\
15 & 3:32:41.18 -27:49:14.84 & 0.00 & 12.6 & 28.99$\pm$ 0.15 &  26.99$\pm$ 0.04 & 1.99  & & \\
16 & 3:32:39.06 -27:45:38.77 & 0.21 &  5.2 & 28.38$\pm$ 0.08 &  27.00$\pm$ 0.04 & 1.38  & & \\
17 & 3:32:36.45 -27:48:34.24 & 0.00 & 14.5 & 28.56$\pm$ 0.12 &  27.12$\pm$ 0.06 & 1.44  & & \\
18 & 3:32:37.28 -27:48:54.58 & 0.16 &  6.8 & 30.36$\pm$ 0.56 &  27.25$\pm$ 0.06 & 3.12  & & \\
19 & 3:32:31.30 -27:48:08.28 & 0.00 & 12.8 & 28.74$\pm$ 0.14 &  27.31$\pm$ 0.06 & 1.44  & & \\
20 & 3:32:33.78 -27:48:07.60 & 0.07 &  6.1 & 29.07$\pm$ 0.10 &  27.35$\pm$ 0.04 & 1.72  & & \\
21 & 3:32:29.45 -27:47:40.52 & 0.00 & 17.2 & 29.34$\pm$ 0.24 &  27.37$\pm$ 0.07 & 1.97  & & \\
22 & 3:32:33.21 -27:46:43.28 & 0.00 & 14.5 & 28.93$\pm$ 0.14 &  27.39$\pm$ 0.06 & 1.54  & & \\
23 & 3:32:38.28 -27:47:51.29 & 0.00 & 11.9 & 28.94$\pm$ 0.14 &  27.50$\pm$ 0.06 & 1.44  & & \\
24 & 3:32:39.86 -27:46:19.09 & 0.00 &  9.9 & 31.31$\pm$ 1.27 &  27.61$\pm$ 0.07 & 3.70  & & \\
25 & 3:32:38.50 -27:48:57.82 & 0.00 & 10.7 & 29.22$\pm$ 0.18 &  27.63$\pm$ 0.08 & 1.58  & & \\
26 & 3:32:44.70 -27:47:11.58 & 0.01 &  9.3 & 29.48$\pm$ 0.23 &  27.65$\pm$ 0.08 & 1.83  & & \\
27 & 3:32:36.62 -27:47:50.03 & 0.00 & 11.4 & 31.89$\pm$ 2.27 &  27.65$\pm$ 0.08 & 4.24  & & \\
28 & 3:32:36.97 -27:45:57.60 & 0.00 & 11.8 &       ---       &  27.72$\pm$ 0.08 & ---   & & \\
29 & 3:32:40.92 -27:48:44.75 & 0.11 &  6.8 & 29.24$\pm$ 0.15 &  27.75$\pm$ 0.07 & 1.49  & & \\
30 & 3:32:33.55 -27:46:44.04 & 0.00 & 13.2 & 29.30$\pm$ 0.20 &  27.78$\pm$ 0.09 & 1.53  & & \\
31 & 3:32:45.16 -27:48:05.11 & 0.00 & 10.2 & 29.21$\pm$ 0.18 &  27.83$\pm$ 0.09 & 1.38  & & \\
32 & 3:32:35.05 -27:47:40.16 & 0.04 &  8.1 & 30.81$\pm$ 0.64 &  27.84$\pm$ 0.07 & 2.97  & & \\
33 & 3:32:42.60 -27:48:08.82 & 0.01 &  7.7 & 29.77$\pm$ 0.23 &  27.86$\pm$ 0.07 & 1.91  & & \\
34 & 3:32:39.45 -27:45:43.42 & 0.45 &  6.2 & 30.98$\pm$ 0.76 &  27.89$\pm$ 0.08 & 3.10  & & \\
35 & 3:32:34.00 -27:48:25.02 & 0.00 &  9.5 & 29.37$\pm$ 0.20 &  27.94$\pm$ 0.09 & 1.43  & & \\
36 & 3:32:38.55 -27:46:17.54 & 0.02 &  7.3 & 29.48$\pm$ 0.16 &  27.97$\pm$ 0.07 & 1.51  & & \\
37 & 3:32:32.36 -27:47:02.83 & 0.00 &  9.6 & 30.03$\pm$ 0.34 &  27.99$\pm$ 0.09 & 2.05  & & \\
38 & 3:32:43.02 -27:46:23.66 & 0.08 &  5.7 &       ---       &  28.00$\pm$ 0.08 & ---   & & \\
{\bf 39a} & 3:32:41.43 -27:46:01.16 & 0.22 &  5.9 &       ---       &  28.00$\pm$ 0.09 & ---   & & \\
{\bf 39b} & 3:32:41.44 -27:46:01.31 & 0.00 &  9.1 &       ---       &  28.14$\pm$ 0.10 & ---   & & \\
40 & 3:32:42.80 -27:48:03.24 & 0.00 &  8.2 &       ---       &  28.06$\pm$ 0.08 & ---   & & \\
41 & 3:32:30.69 -27:46:54.84 & 0.92 &  4.0 & 29.72$\pm$ 0.33 &  28.10$\pm$ 0.13 & 1.62  & & \\
42 & 3:32:35.08 -27:48:06.80 & 0.09 &  6.8 & 29.98$\pm$ 0.29 &  28.13$\pm$ 0.09 & 1.85  & & \\
43 & 3:32:37.23 -27:45:38.38 & 0.00 & 10.9 &       ---       &  28.14$\pm$ 0.09 & ---   & & \\
44 & 3:32:38.79 -27:47:10.86 & 0.00 &  9.8 & 29.48$\pm$ 0.19 &  28.16$\pm$ 0.10 & 1.32  & & \\
45 & 3:32:40.06 -27:49:07.50 & 0.94 &  4.0 & 29.97$\pm$ 0.16 &  28.20$\pm$ 0.05 & 1.77  & & \\
46 & 3:32:44.70 -27:46:45.44 & 0.47 &  4.5 & 30.06$\pm$ 0.24 &  28.27$\pm$ 0.08 & 1.79  & & \\
47 & 3:32:44.14 -27:48:27.07 & 0.61 &  4.6 & 30.17$\pm$ 0.29 &  28.30$\pm$ 0.09 & 1.86  & & \\
48 & 3:32:34.14 -27:48:24.37 & 0.60 &  5.3 & 30.54$\pm$ 0.44 &  28.38$\pm$ 0.11 & 2.16  & & \\
49 & 3:32:42.20 -27:49:12.00 & 0.52 &  7.8 & 31.48$\pm$ 1.21 &  28.39$\pm$ 0.12 & 3.09  & & \\
50 & 3:32:43.16 -27:48:29.56 & 0.58 &  7.1 & 29.75$\pm$ 0.21 &  28.43$\pm$ 0.11 & 1.32  & & \\
51 & 3:32:34.58 -27:46:58.01 & 0.13 &  5.7 & 31.49$\pm$ 1.00 &  28.44$\pm$ 0.11 & 3.05  & & \\
52 & 3:32:40.25 -27:46:05.12 & 0.74 &  5.0 & 31.30$\pm$ 0.88 &  28.44$\pm$ 0.11 & 2.86  & & \\
53 & 3:32:39.48 -27:48:40.10 & 0.48 &  7.9 & 31.55$\pm$ 1.10 &  28.46$\pm$ 0.11 & 3.10  & & \\
54 & 3:32:39.52 -27:45:13.39 & 0.90 &  5.0 &       ---       &  28.49$\pm$ 0.15 & ---   & & \\
55 & 3:32:33.33 -27:48:24.16 & 0.00 &  7.4 & 30.82$\pm$ 0.54 &  28.50$\pm$ 0.11 & 2.32  & & \\
{\bf 56a} & 3:32:34.52 -27:47:34.84 & 0.10 &  5.5 & 31.16$\pm$ 0.70 &  28.52$\pm$ 0.11 & 2.65  & & \\
{\bf 56b} & 3:32:34.47 -27:47:35.05 & 0.25 &  6.4 & 30.44$\pm$ 0.37 &  28.65$\pm$ 0.12 & 1.79  & & \\
57 & 3:32:40.56 -27:48:02.59 & 0.00 &  7.9 &       ---       &  28.54$\pm$ 0.12 & ---   & & \\
58 & 3:32:39.41 -27:47:59.42 & 0.01 &  7.9 & 30.29$\pm$ 0.31 &  28.58$\pm$ 0.11 & 1.71  & & \\
59 & 3:32:38.86 -27:47:13.16 & 0.88 &  3.9 & 30.10$\pm$ 0.20 &  28.61$\pm$ 0.09 & 1.49  & & \\
60 & 3:32:47.56 -27:47:11.33 & 0.73 &  4.2 & 30.18$\pm$ 0.29 &  28.62$\pm$ 0.12 & 1.56  & & \\
61 & 3:32:32.72 -27:46:37.24 & 0.49 &  5.1 & 30.24$\pm$ 0.30 &  28.63$\pm$ 0.12 & 1.61  & & \\
62 & 3:32:47.97 -27:47:05.14 & 0.22 &  5.0 &       ---       &  28.65$\pm$ 0.14 & ---   & & \\
63 & 3:32:35.10 -27:48:09.14 & 0.00 &  9.6 & 30.04$\pm$ 0.22 &  28.68$\pm$ 0.11 & 1.36  & & \\
64 & 3:32:40.82 -27:47:48.77 & 0.75 &  6.4 & 30.62$\pm$ 0.36 &  28.81$\pm$ 0.12 & 1.81  & & \\
65 & 3:32:40.26 -27:48:08.10 & 0.86 &  3.7 &       ---       &  28.82$\pm$ 0.09 & ---   & & \\
66 & 3:32:40.53 -27:45:46.48 & 0.66 &  4.7 & 30.43$\pm$ 0.28 &  28.83$\pm$ 0.11 & 1.60  & & \\
67 & 3:32:46.17 -27:47:45.31 & 0.18 &  5.2 & 31.28$\pm$ 0.64 &  28.84$\pm$ 0.12 & 2.44  & & \\
68 & 3:32:38.05 -27:45:48.82 & 0.10 &  7.1 &       ---       &  28.85$\pm$ 0.14 & ---   & & \\
69 & 3:32:37.69 -27:46:21.54 & 0.71 &  6.9 & 33.74$\pm$ 7.51 &  28.86$\pm$ 0.15 & 4.88  & & \\
70 & 3:32:36.77 -27:48:56.95 & 0.77 &  7.5 & 30.24$\pm$ 0.30 &  28.86$\pm$ 0.15 & 1.38  & & \\
71 & 3:32:42.71 -27:48:11.81 & 0.31 &  6.2 & 30.91$\pm$ 0.49 &  28.87$\pm$ 0.13 & 2.04  & & \\
72 & 3:32:39.13 -27:48:18.47 & 0.58 &  5.4 & 31.83$\pm$ 1.16 &  28.90$\pm$ 0.14 & 2.94  & & \\
73 & 3:32:38.16 -27:47:33.36 & 0.11 &  6.3 & 31.07$\pm$ 0.68 &  28.91$\pm$ 0.16 & 2.16  & & \\
74 & 3:32:45.33 -27:47:03.52 & 0.71 &  4.0 & 31.78$\pm$ 1.09 &  28.94$\pm$ 0.14 & 2.84  & & \\
75 & 3:32:36.77 -27:47:53.59 & 0.65 &  6.1 &       ---       &  28.96$\pm$ 0.16 & ---   & & \\
76 & 3:32:42.78 -27:46:37.96 & 0.82 &  4.0 & 30.33$\pm$ 0.28 &  28.97$\pm$ 0.14 & 1.36  & & \\
77 & 3:32:40.13 -27:49:36.98 & 0.08 &  5.8 & 32.41$\pm$ 2.11 &  28.97$\pm$ 0.16 & 3.43  & & \\
78 & 3:32:34.09 -27:47:57.55 & 0.11 &  6.6 & 30.34$\pm$ 0.26 &  28.99$\pm$ 0.13 & 1.35  & & \\
79 & 3:32:30.55 -27:47:16.69 & 0.81 &  4.5 &       ---       &  29.00$\pm$ 0.14 & ---   & & \\
80 & 3:32:44.98 -27:46:40.69 & 0.02 &  6.7 &       ---       &  29.00$\pm$ 0.15 & ---   & & \\
81 & 3:32:37.83 -27:49:05.84 & 0.75 &  5.3 & 31.34$\pm$ 0.63 &  29.01$\pm$ 0.13 & 2.33  & & \\
82 & 3:32:39.79 -27:46:33.74 & 0.02 &  7.4 & 31.27$\pm$ 0.47 &  29.03$\pm$ 0.11 & 2.24  & & \\
83 & 3:32:38.27 -27:46:18.44 & 0.39 &  2.1 & 31.22$\pm$ 0.49 &  29.22$\pm$ 0.14 & 2.00  & & \\
84 & 3:32:44.49 -27:47:13.74 & 0.61 &  3.0 & 32.28$\pm$ 1.39 &  29.22$\pm$ 0.15 & 3.06  & & \\
85 & 3:32:42.19 -27:46:27.88 & 0.78 &  6.1 &       ---       &  29.23$\pm$ 0.16 & ---   & & \\
86 & 3:32:41.18 -27:46:55.24 & 0.75 &  3.5 & 31.03$\pm$ 0.47 &  29.25$\pm$ 0.16 & 1.78  & & \\
87 & 3:32:31.55 -27:48:13.97 & 0.29 &  4.3 & 32.88$\pm$ 2.38 &  29.31$\pm$ 0.15 & 3.58  & & \\
88 & 3:32:40.44 -27:46:32.12 & 0.59 &  4.5 & 31.92$\pm$ 1.02 &  29.32$\pm$ 0.16 & 2.59  & & \\
89 & 3:32:39.50 -27:46:49.44 & 0.70 &  4.1 & 32.28$\pm$ 1.35 &  29.33$\pm$ 0.16 & 2.95  & & \\
90 & 3:32:40.83 -27:48:31.64 & 0.50 &  5.3 & 31.07$\pm$ 0.44 &  29.36$\pm$ 0.16 & 1.72  & & \\
91 & 3:32:35.04 -27:47:25.76 & 0.73 &  5.7 & 30.94$\pm$ 0.41 &  29.36$\pm$ 0.17 & 1.58  & & \\
92 & 3:32:47.72 -27:47:02.58 & 0.53 &  3.5 & 30.86$\pm$ 0.39 &  29.37$\pm$ 0.17 & 1.48  & & \\
93 & 3:32:35.90 -27:49:02.75 & 0.79 &  5.2 & 30.97$\pm$ 0.40 &  29.40$\pm$ 0.17 & 1.57  & & \\
94 & 3:32:39.24 -27:49:23.74 & 0.47 &  3.9 & 30.91$\pm$ 0.34 &  29.41$\pm$ 0.15 & 1.51  & & \\
95 & 3:32:43.79 -27:46:33.71 & 0.41 &  6.1 & 30.85$\pm$ 0.34 &  29.42$\pm$ 0.16 & 1.43  & & \\
96 & 3:32:38.68 -27:49:12.86 & 0.63 &  2.4 & 31.38$\pm$ 0.53 &  29.49$\pm$ 0.16 & 1.89  & & \\
97 & 3:32:41.33 -27:49:20.46 & 0.63 &  4.1 & 31.19$\pm$ 0.50 &  29.52$\pm$ 0.19 & 1.66  & & \\
98 & 3:32:39.42 -27:47:02.58 & 0.53 &  3.4 & 31.82$\pm$ 0.74 &  29.62$\pm$ 0.17 & 2.20  & & \\
99 & 3:32:47.81 -27:47:20.54 & 0.69 &  4.1 & 31.29$\pm$ 0.47 &  29.62$\pm$ 0.17 & 1.68  & & \\
100 & 3:32:31.75 -27:46:50.48 & 0.67 & 4.3 &       ---       &  29.65$\pm$ 0.20 & ---   & & \\
101 & 3:32:34.19 -27:46:39.07 & 0.57 & 3.7 &       ---       &  29.97$\pm$ 0.20 & ---   & & \\

\enddata
 
\tablenotetext{1.} {Photometric properties of the 108 $z\simeq 6$ candidates
discovered in the UDF. The first 12 objects are the candidates that have been
identified in the UDF NICMOS images. Only these 12 objects appear in print,
and the full table is available in the electronic version.}
\tablenotetext{2.} {There are six multiple systems among these candidates, four
of which are within the NICMOS UDF area. The members of these systems have
their ID in bold face. Note that object 1a is a known galaxy at $z=5.83$.}
\tablenotetext{3.} {S/G is star/galaxy separation code, with 0 
for the most extended sources and 1 for the most compact sources.}
\tablenotetext{4.} {FWHM (in pixel) is derived by assuming a Guassian core.}
\end{deluxetable}